\PassOptionsToPackage{table}{xcolor}
\documentclass[sigconf]{acmart}

\AtBeginDocument{%
  }

\usepackage[english]{babel}
\usepackage[utf8]{inputenc}

\usepackage{flushend}
\usepackage{balance}
\usepackage{multirow}
\usepackage{multirow}

\usepackage{color}
\usepackage{booktabs}
\usepackage{fancybox}
\usepackage{multirow}
\usepackage{float}
\usepackage{array,graphicx}
\usepackage{subcaption}
\usepackage[flushleft]{threeparttable}
\usepackage[most]{tcolorbox}
\usepackage{xspace}
\usepackage{listings}
\usepackage{pifont}
\usepackage{tikz}
\usepackage{blindtext}
\definecolor{purplish}{HTML}{D8DFE3}
\definecolor{purplishlight}{HTML}{EBEFF3}
\definecolor{purplishdark}{HTML}{009e73}

\newtcolorbox[auto counter,number within=section]{rqbox}[2]{
    nameref=#1,
    title=\small{#1}, 
    enhanced,
    attach boxed title to top left={yshift=-6pt, xshift=8pt},
    boxed title style={size=small,boxsep=1pt},
    colframe=purplishdark,colback=white,colbacktitle=purplishdark,
    boxsep=2pt,left=2pt,right=2pt,top=6pt,bottom=2pt,middle=2pt
}

\newcommand{\rqtextone}{What factors are associated with students' interest in contributing to an OSS project?}

\newcommand{\rqtexttwo}{What recommendations do students have to make OSS projects more attractive?}

\copyrightyear{2025}
\acmYear{2025}
\setcopyright{cc}
\setcctype{by}
\acmConference[FSE Companion '25]{33rd ACM International Conference on the Foundations of Software Engineering}{June 23--28, 2025}{Trondheim, Norway}
\acmBooktitle{33rd ACM International Conference on the Foundations of Software Engineering (FSE Companion '25), June 23--28, 2025, Trondheim,
Norway}\acmDOI{10.1145/3696630.3727243}
\acmISBN{979-8-4007-1276-0/2025/06}

\balance

\begin{document}

\title[Exploring the Untapped: Student Perceptions and Participation in OSS]{Exploring the Untapped: Student Perceptions \\and Participation in OSS}

\author{Italo Santos}
\email{italo.santos@nau.edu}
\affiliation{%
  \institution{Northern Arizona University \\ University of Hawai‘i at Mānoa}
  \city{Flagstaff}
  \state{AZ}
  \country{USA}
}

\author{Katia Romero Felizardo}
\email{katiascannavino@utfpr.edu.br}
\affiliation{%
  \institution{Federal University of Technology}
  \city{Paraná}
  \state{PR}
  \country{Brazil}
}

\author{Bianca Trinkereinch}
\email{bianca.trinkenreich@colostate.edu}
\affiliation{%
  \institution{Colorado State University}
  \city{Fort Collins}
  \state{CO}
  \country{USA}
}

\author{Daniel M. German}
\email{dmg@uvic.ca}
\affiliation{%
  \institution{University of Victoria}
  \city{Victoria}
  %\state{XX}
  \country{Canada}
}

\author{Igor Steinmacher}
\email{igor.steinmacher@nau.edu}
\affiliation{%
  \institution{Northern Arizona University}
  \city{Flagstaff}
  \state{AZ}
  \country{USA}
}

\author{Marco A. Gerosa}
\email{marco.gerosa@nau.edu}
\affiliation{%
  \institution{Northern Arizona University}
  \city{Flagstaff}
  \state{AZ}
  \country{USA}
}

\renewcommand{\shortauthors}{Santos et al.}

\begin{abstract}
    Open Source Software (OSS) projects offer valuable opportunities to train the next generation of software engineers while benefiting projects and society as a whole. While research has extensively explored student participation in OSS and its use in software engineering education, student participation in OSS is still low, and the perspectives of students who have never contributed remain underexplored. This study aims to investigate the relationship between students' interest in contributing to OSS and their perceptions of barriers and motivational factors. We developed a theoretical model to understand the relationship between students’ perceptions of OSS and their interest in contributing. We then surveyed students majoring in computer science and related fields (N=241). Using structural equation modeling techniques, we tested the model and found that intrinsic and internalized extrinsic motivations are positively associated with interest in contributing to OSS projects, while the impact of extrinsic motivation varies by gender. Comparatively, we found no significant relationship between barriers and interest in contributing. Students suggested several ways to make projects more attractive, including increasing awareness of the importance of OSS. Our findings can help communities better prepare to integrate students and encourage educators to enhance interest in OSS by linking participation to specific motivational factors.
\end{abstract}





\maketitle

%\vspace{-5px}
\section{Introduction}
\label{sec:introduction}
%\vspace{-2px}

Fostering students' engagement in open source software (OSS) as part of Computer Science courses helps to train the future software engineering workforce~\cite{silva2020google, pinto2019training, pinto2017training}. Contributing to OSS enables individuals starting their careers in software development to acquire technical and soft skills in practice. Participation in OSS can significantly boost confidence when seeking industry positions~\cite{pinto2019training, braught2018multi, morgan2014lessons, nascimento2013using}, providing real-world experience, enhancing skills, and expanding professional networks. Additionally, successful participation in OSS projects increases students' visibility among their peers~\cite{cai2016reputation, riehle2015open} and is considered by major tech companies in hiring processes~\cite{dekoenigsberg2008successful}. Beyond individual and project-level advantages, OSS contributions also benefit society by further developing widely used products~\cite{parra2016making} and strengthening the future workforce~\cite{greene2016cvexplorer, riehle2015open}.

However, those in the initial stages of their career, such as students, often do not participate in OSS. Students have characteristics that make them particularly suitable for participating in OSS projects~\cite{ellis2008support,steinmacher2015social}, and OSS projects offer a variety of tasks, including some that do not involve coding. Moreover, since OSS projects vary considerably in domain, size, and complexity~\cite{von2003special, sarma2016training}, they offer a wide spectrum of choices. Nonetheless, students, especially those from underrepresented populations, often feel demotivated to participate due to barriers that hinder contributions~\cite{steinmacher2015social, padala2020gender, trinkenreich2021women} or for not understanding how OSS can be motivating or useful for their goals.

Previous work focused on attracting developers~\cite{steinmacher2014attracting, hannebauer2017relationship, zhang2019companies, butler2019company, pinto2019training}, the motivations of current developers~\cite{gerosa2021shifting, von2012carrots}, and why students participate in OSS summer programs~\cite{FSE_GSoC}. Understanding the factors associated with student interest in contributing helps develop effective attraction strategies. Student motivations, barriers, and perceptions may differ from those of more experienced contributors or those already in the job market. %This study provides insights into how to engage this group. Contributing is an effective way to develop practical skills, gain real-life experience, and boost employability~\cite{pinto2017training, silva2020google}. Students are an important population to study due to their unique characteristics like age, experience, and background.
Unlike existing literature, which considers only students already involved in OSS~\cite{pinto2017training,pinto2019training,nascimento2019does,FSE_GSoC} and leads to a survivability bias, our study, to the best of our knowledge, is the first to include students who have never contributed to OSS, providing new insights into this group. Understanding the perception of those who have never contributed is relevant since they represent potential future contributors, and their views can evidence barriers to entry or misconceptions about OSS that projects, and educators may need to address or benefits they can highlight.

%Our study is the first, to the best of our knowledge, to include students who have never contributed to OSS, providing new insights into this group. Unlike existing literature, which considers only students already involved in OSS~\cite{pinto2017training,pinto2019training,nascimento2019does,FSE_GSoC}, leading to a survivability bias, our investigation includes those who have never participated. %Specifically, we understand how the perceptions about OSS motivational factors and barriers correlate with interest in contributing. We are not focusing on current practices but on how students perceive OSS. Thus, 
%Understanding the perception of those who have never contributed is relevant since they represent potential future contributors, and their views can evidence barriers to entry or misconceptions about OSS that projects may need to address or benefits they can highlight.}

The aim of our study is twofold. First, we seek to understand how different factors (perceptions of barriers and motivation) are associated with interest in contributing to OSS. Second, we aim to collect student recommendations on making OSS projects more attractive. To guide our investigation, we pose the following research questions: 

%We pose the  research questions: \textbf{RQ1:} \textit{\rqtextone} and \textbf{RQ2:} \textit{\rqtexttwo}

 \newcommand{\rqone}[2][]{
     \begin{rqbox}{\textbf{Research Question 1}}{#2}
         \rqtextone
         #1
     \end{rqbox}
 }

 \rqone{}

 \newcommand{\rqtwo}[2][]{
     \begin{rqbox}{\textbf{Research Question 2}}{#2}
         \rqtexttwo
         #1
     \end{rqbox}
 }

 \rqtwo{}

We developed a theoretical model of interest to contribute to OSS grounded in prior literature (Sec.~\ref{sec:theorydevelopment}). We evaluated our model through a survey (N=241) of students majoring in computer science and related fields (e.g., software engineering, information systems, and computer engineering) using partial least squares structural equation modeling (PLS-SEM) (Sec.~\ref{sec:methodology}). Our analysis provided empirical support for part of our model, showing that the factors associated with students' interest in contributing to OSS projects include intrinsic and internalized extrinsic motivations positively associated with interest in OSS (Sec.~\ref{sec:results}). %Furthermore, we found that the impact of extrinsic motivation varies by gender and that the student's previous experience is associated with the interest in contributing to OSS. We did not find a significant association between barriers to contribute and interest. 

Among our contributions, our results reveal strategies to make OSS projects more attractive to students. We found that fun is a strong component of intrinsic motivation, which is the most correlated factor to interest in contributing. %Therefore, projects and educators can focus on advertising how contributing can be fun (e.g., testimonies) instead of clarifying how projects deal with specific barriers (e.g., welcomeness) since barriers are not the most determinant factors for the interest in contributing. 
Moreover, we identified gender differences in the association of extrinsic motivation and interest, which has consequences for communities and educators aiming at promoting diversity in OSS. We also brought to the foreground the voices of students who suggested various strategies to make projects more appealing. These included simplifying the contribution process, raising awareness, and offering academic and career-related benefits. Additional strategies involved providing mentorship, recognizing contributors with rewards, fostering a welcoming environment, and aligning projects with individuals' personal interests. Our insights can help educators, project leaders, contributors, and the research community create future interventions to attract students to OSS projects.

\vspace{-5px}
\section{Theory Development}
\label{sec:theorydevelopment}
\vspace{-2px}

Previous studies have identified several barriers that influence newcomers’ onboarding experiences~\cite{wolff2013patterns,steinmacher2015understanding}. When faced with barriers, newcomers often lose motivation and give up~\cite{balali2020recommending, steinmacher2015understanding, wang2011bug}.
This particularly impacts newcomer students, who typically have a limited skill set and experience when first contributing to an OSS project. 
%For example, toxic environments have been highlighted in the literature, where OSS project members have been found to be unfriendly, unhelpful, or behave as elitists~\cite{bosu2019diversity, mall2020four, paul2019expressions, storey2016social}. Steimacher et al.~\cite{steinmacher2019overcoming} notes that newcomers are sometimes unaware of community communication protocols~\cite{jensen2011joining, steinmacher2013newcomers}. %Jensen et al.~\cite{jensen2011joining} analyzed the speed at which emails sent by newcomers are answered and the impact of gender or nationality on the responses and reception they receive.
Balali et al.~\cite{balali2018newcomers}, for example, identified barriers related to low self-efficacy, where newcomers believe that they will be unable to complete assigned tasks. Yu et al.~\cite{yu2012empirical} found a set of process barriers that impact newcomers. Technical complexity may also scare newcomers~\cite{steinmacher2019overcoming}. Steinmacher et al.~\cite{steinmacher2016overcoming} reported that the time required to contribute often leads to frustration or demotivation among newcomers. 
%These feelings typically arise during activities with unsatisfactory outcomes, indicating the significant impact of time-consuming tasks on newcomers' motivation. 
Hence, we hypothesize that: 

%\begin{itemize}
%\item 
\textbf{Hypothesis H1:} Students who believe that OSS poses onboarding barriers are less interested in contributing to OSS projects.
%\end{itemize}

Many researchers have investigated the motivations to contribute to OSS~\cite{lakhani2004open, bitzer2007intrinsic, von2012carrots, gerosa2021shifting}. Von Krogh et al.~\cite{von2012carrots} surveyed the literature and identified ten motivations grouped into three main types: (i) extrinsic, (ii) internalized extrinsic, and (iii) intrinsic. Individuals are extrinsically motivated when they seek external incentives or alter their behavior due to external interventions~\cite{frey1997relationship}. Extrinsic motivations include career advancement and pay. In contrast, developers can also internalize extrinsic motivators, perceiving them as self-regulating behaviors rather than external impositions~\cite{deci1987support}. These internalized extrinsic motivations include reputation, reciprocity, learning, and own use~\cite{von2012carrots}. Intrinsic motivations, on the other hand, are those that occur when an action is performed for the inherent joy of performing it rather than in response to external pressures or rewards~\cite{ryan2000self} and include ideology, altruism, kinship, and fun~\cite{von2012carrots}. Gerosa et al.~\cite{gerosa2021shifting} found that intrinsic and internalized extrinsic motivations explain what drives most contributors. Other studies suggest that extrinsic motivation, such as future monetary rewards, motivates developers~\cite{alexander2002working}. Thus, we define the following hypotheses:  

%\begin{itemize}
    %\item 
\textbf{Hypothesis H2:} Students who perceive extrinsic motivational factors in the OSS environment are more interested in contributing to OSS projects. 

    %\item  
\textbf{Hypothesis H3:} Students who perceive internalized extrinsic motivational factors in the OSS environment are more interested in contributing to OSS projects.

    %\item 
\textbf{Hypothesis H4:} Students who perceive intrinsic motivational factors in the OSS environment are more interested in contributing to OSS projects.

%\end{itemize}
Low diversity in OSS is a concern raised by various studies. In this study, we focus on gender diversity since it is well known that gender minorities face challenges in becoming part of the OSS community~\cite{trinkenreich2021women, guizani2022debug}, which discourages their participation~\cite{miller2012toward, vugt2007gender}. Understanding what is associated with different populations can help better support diversity. Hence, we propose the following four moderating hypotheses: 

%\begin{itemize}
    %\item 
\textbf{Hypothesis H5a:} Gender minorities are less interested in contributing when they believe that OSS projects pose onboarding barriers. 

    %\item 
\textbf{Hypothesis H5b:} Gender minorities are less interested in contributing when they perceive extrinsic motivational factors in OSS. 

    % \item 
\textbf{Hypothesis H5c:} Gender minorities are less interested in contributing when they perceive internalized extrinsic motivational factors in OSS. 

    %\item 
\textbf{Hypothesis H5d:} Gender minorities are less interested in contributing when they perceive intrinsic motivational factors in OSS. 
%\end{itemize}

In addition, we used the control variables \textit{OSSCourse} and \textit{Involvement with OSS}. % to ensure a more accurate understanding of the factors associated with students' interests.
\textit{OSSCourse} refers to whether the student has taken an OSS course. Including this variable helps to account for the influence that formal education and exposure to OSS concepts in an academic setting might have on the student's interest in contributing to OSS projects. It controls for the additional knowledge and motivation that might arise from structured learning environments. The control variable \textit{Involvement with OSS} categorizes students into three types: (i) never contributed, (ii) dropped out, and (iii) contributor. This categorization helps control for the students' varying levels of experience and engagement with OSS, assessing how prior involvement impacts students' interests.

%\\
%\begin{itemize}
    %\item 
%\textit{OSSCourse}: This control variable refers to whether the student has taken an OSS course. Including this variable helps to account for the influence that formal education and exposure to OSS concepts in an academic setting might have on the student's interest in contributing to OSS projects. It controls for the additional knowledge and motivation that might arise from structured learning environments.
    %\item
%\end{itemize}

\begin{table*}[!ht]\scriptsize
\centering
\caption{Constructs and their corresponding survey questions}
\label{tab:surveyquestions}
\vspace{-4px}
\begin{tabular}{l|l|l}
\hline
\textbf{Construct} & \textbf{Category} & \textbf{Item question} \\ \hline \hline

\multirow{7}{*}{Barriers} & Toxicity & OSS is a toxic environment~\textsuperscript{1, 5} \\ \cline{2-3} 

 & Communication issues & OSS is an environment in which it is difficult to communicate with other members~\textsuperscript{1, 3, 5} \\ \cline{2-3} 
 & Welcomeness & OSS is a welcoming environment for external contributors~\textsuperscript{1, 3, 5} \textit{(reverse coded)}\\ \cline{2-3} 
 & Confusing process & It is difficult to understand how to contribute to OSS~\textsuperscript{1, 3, 5} \\ \cline{2-3} 
 & Low self-efficacy & OSS is only for smart people~\textsuperscript{3, 5} \\ \cline{2-3} 
 & Time-consuming & Contributing to OSS is too time-consuming~\textsuperscript{3, 5} \\ \cline{2-3} 
 & Technical complexity & OSS is too technical and complex~\textsuperscript{1, 3, 5} \\ \hline \hline
 
\multirow{2}{*}{Extrinsic} & Career & Contributing to OSS increases job opportunities~\textsuperscript{2, 4, 6, 7} \\ \cline{2-3} 
 & Pay & OSS is an environment in which it is possible to earn money~\textsuperscript{2, 7} \\ \hline \hline
 
\multirow{4}{*}{\begin{tabular}[c]{@{}l@{}}Internalized \\ extrinsic\end{tabular}} & Reputation & Contributing to OSS can enhance developers' reputation~\textsuperscript{2, 4, 6, 7} \\ \cline{2-3} 
 & Own-use & OSS can be for working on software one needs for their professional or personal purposes~\textsuperscript{2, 4, 6} \\ \cline{2-3} 
 & Learning & OSS is a place in which people learn and improve their skills~\textsuperscript{2, 4, 6, 7} \\ \cline{2-3} 
 & Reciprocity & OSS members feel personally obligated to contribute because they use OSS products~\textsuperscript{2, 6} \\ \hline \hline
 
\multirow{5}{*}{Intrinsic} & Fun & OSS projects are fun environments~\textsuperscript{2, 6} \\ \cline{2-3} 
 & Kinship & OSS is a collaborative environment~\textsuperscript{2, 4, 7} \\ \cline{2-3} 
 & \begin{tabular}[c]{@{}l@{}}Altruism (share knowledge)\end{tabular} & OSS is an environment in which people love sharing knowledge and skills and helping others~\textsuperscript{2, 7} \\ \cline{2-3} 
 & \begin{tabular}[c]{@{}l@{}}Altruism (benefit society)\end{tabular} & OSS provides software products that benefit society~\textsuperscript{2, 4} \\ \cline{2-3} 
 & Ideology & \begin{tabular}[c]{@{}l@{}}OSS software enables knowledge to be open and to limit the power of proprietary software \\ and large companies~\textsuperscript{2, 6, 7}\end{tabular} \\ \hline \hline

 \multicolumn{2}{l|}{Interest to Contribute to OSS} & \multicolumn{1}{l}{How do you rate your interest in contributing to an OSS project?} \\ \hline \hline

 \multicolumn{3}{l}{\begin{tabular}[c]{@{}l@{}}

 \textsuperscript{1}~\scriptsize{Adapted from Steinmacher et al.~\cite{steinmacher2015social}}, 

 \textsuperscript{2}~\scriptsize{Adapted from Gerosa et al.~\cite{gerosa2021shifting}},
 
 \textsuperscript{3}~\scriptsize{Adapted from Balali et al.~\cite{balali2018newcomers}}, 
 
 \textsuperscript{4}~\scriptsize{Adapted from Hars and Ou~\cite{alexander2002working}}\\
 
 \textsuperscript{5}~\scriptsize{Adapted from Guizani et al.~\cite{guizani2021long}}, 
 
 \textsuperscript{6}~\scriptsize{Adapted from Lakhani and Wolf~\cite{lakhani2005hackers}},
 
 \textsuperscript{7}~\scriptsize{Adapted from Ghosh et al.~\cite{ghosh2002free}}

 \end{tabular}} \\ \hline \hline
 
\end{tabular}
\vspace{-10px}
\end{table*}

\vspace{-5px}
\section{Research Design}
\label{sec:methodology}
\vspace{-2px}

We surveyed students majoring in computer science and related fields (e.g., software engineering, information systems, and computer engineering). Surveys are suitable for gathering numerous responses necessary to evaluate a theoretical model such as ours~\cite{barcomb2019episodic}. Then, we used the Partial Least Squares Structural Equation Modeling (PLS-SEM) to analyze our survey data and explain the variances of dependent variables. PLS-SEM's predictive capabilities~\cite{hair2021primer} align well with our research objectives since it performs well with a limited number of indicators per construct, handles formative constructs, offers strong statistical power, and is well-suited for complex models comprising multiple constructs and indicators, allowing us to capture the multifaceted nature of our research questions. 
%Additionally, PLS-SEM boasts strong statistical power, enabling the identification of significant relationships within the population~\cite{hair2021primer}. These attributes make PLS-SEM a good choice for our study, ensuring robust and reliable analysis of the relationships between the constructs under investigation. 
%The research design %is summarized in Figure~\ref{fig:researchdesign}. In the following subsections, we discuss the 
%including survey design, participant recruitment, data collection and analysis procedures, and the measurement model (i.e., operationalization of constructs) are explained in the following subsections.

\vspace{-2px}
\subsection{Measurement model}
%\vspace{-2px}

%The hypotheses' theoretical model is based on established OSS literature; some concepts may be directly observed (e.g., ``Involvement with OSS"), while others cannot (e.g., interest in contributing to OSS projects). These concepts are represented as 

Some constructs in the theoretical model are represented by latent variables, as observed in Table~\ref{tab:surveyquestions}. A latent variable cannot be directly measured or observed but is measured through a set of indicators or manifest variables~\cite{trinkenreich2023belong}. For the latent variables in this study, we adapted existing measurement instruments as much as possible to improve construct validity~\cite{von2012carrots, steinmacher2015social, gerosa2021shifting}. We adapted existing measurement instruments applied to the broad OSS literature (including novices and students), iteratively refining them to be clearer for our intended context, e.g., rather than using ``I have fun writing programs'', we used ``OSS projects are fun environments'' to more accurately reflect the broader perception of contributing to OSS. We conducted pilots to gather feedback and ensure the adequate interpretation of the questions by our target population.
%In our study, we have the following constructs: \textbf{Motivation} to contribute to OSS, \textbf{Barriers} faced when trying to contribute to OSS, and \textbf{Interest to contribute} to OSS projects. Gender, OSSCourse, and Involvement with OSS were directly asked. 
%Our survey used validated scales. A construct is defined by a quantity that cannot be measured directly.

\begin{figure}[!ht]
    \centering
    \includegraphics[width=8.5cm]{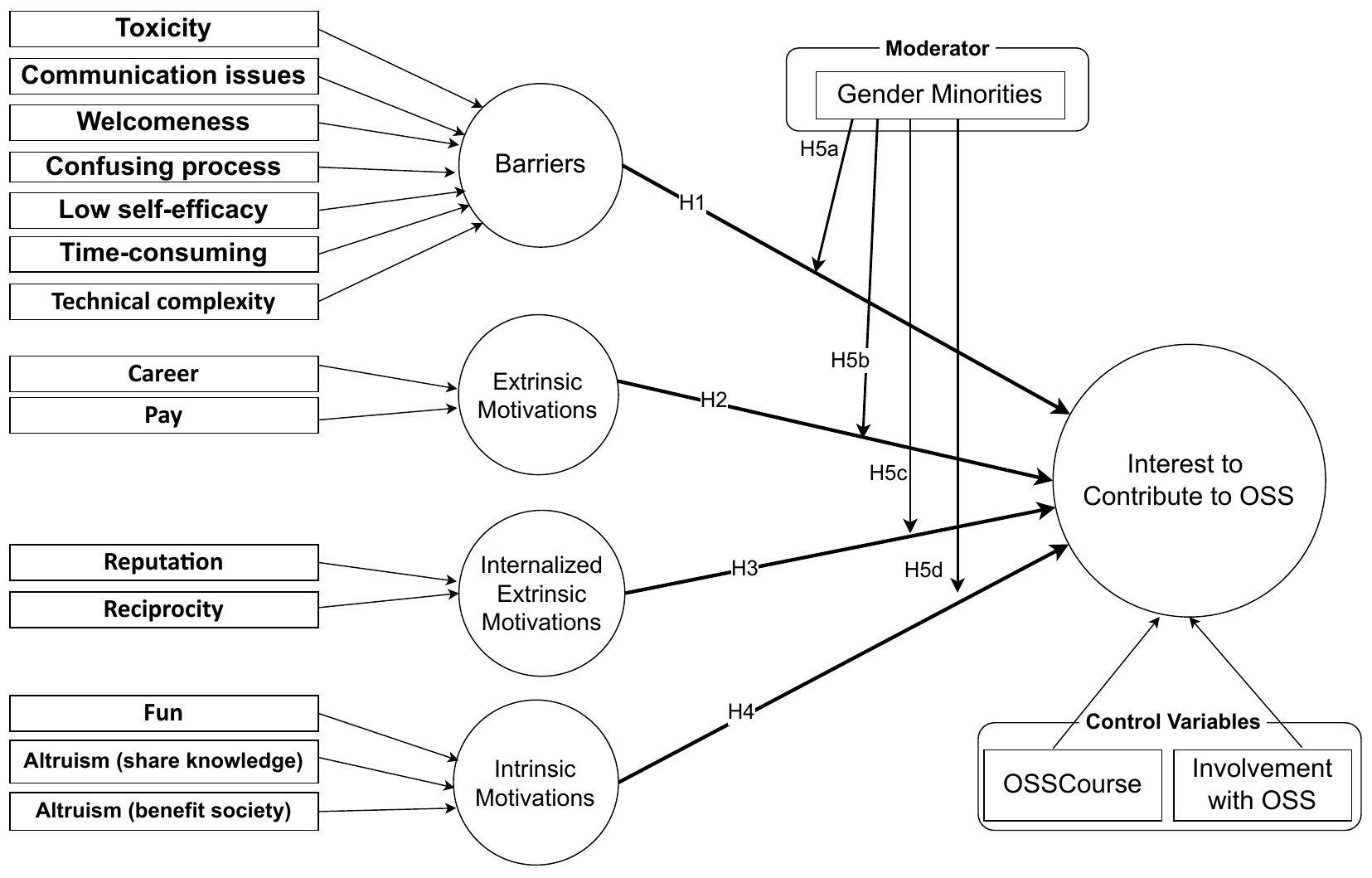}
    \caption{Conceptualized PLS-SEM model.}
    \label{fig:firstmodel}
    \vspace{-15px}
\end{figure}

We selected our constructs based on hypotheses we derived from the literature. %We adapted previous instruments to collect the data according to our context. 
We wanted to investigate students' interest, which can lead to an intention to act~\cite{HONG2014110}. Intention may involve temporal and personal impediments that are outside the control of the projects and educators. The constructs in our model are ``formative''~\cite{hair2021primer}, meaning that every indicator captures a specific aspect of the latent variable; in other words, the indicators are not interchangeable~\cite{hair2021primer}. Our model is illustrated in Figure~\ref{fig:firstmodel} and detailed in the following.

%\begin{itemize}
    %\item 
\textbf{Motivation:} The literature review by Von Krogh et al.~\cite{von2012carrots} is a comprehensive investigation of motivation in OSS since they aggregated motivation factors found in 40 primary studies published until 2009. The authors grouped the motivation factors into ten main categories, namely, \textit{Ideology, Altruism, Kinship, Fun, Reputation, Reciprocity, Learning, Own-Use, Career, and Pay}, which were organized into extrinsic, internalized extrinsic, and intrinsic motivations. We adapted questions from previous empirical studies~\cite{gerosa2021shifting, alexander2002working, lakhani2005hackers, ghosh2002free} that investigated motivations for participating in OSS projects. In our study, respondents self-assessed their perception of the presence of motivational factors to join OSS on a 5-point Likert scale from 1 (strongly disagree) to 5 (strongly agree).

    %\item 
\textbf{Barriers:} Steinmacher et al.~\cite{steinmacher2015social} identified barriers newcomers face during their first attempt to contribute to OSS projects. Those barriers emerged from a literature review on OSS newcomers and interviews and questionnaires conducted with developers in different stages of the OSS community joining process. We adapted our questions from previous empirical studies~\cite{steinmacher2015social, balali2018newcomers, guizani2021long} that examined barriers in OSS projects. 
%We note that some items (e.g., welcomeness) were inverted during the analysis to ensure all responses were on the same scale. 
Our respondents self-assessed the barriers they perceive in OSS on a 5-point Likert scale from 1 (strongly disagree) to 5 (strongly agree).

    %\item
\textbf{Interest to contribute:} Previous studies define interest as a dimension that accounts for performance at individual level~\cite{blumberg1982missing}. Our respondents self-assessed their interest in contributing on a scale of 0 (not interested) to 10 (very interested).
%\end{itemize}

\textbf{Gender:} We adapted questions from surveys used in OSS communities to ask about gender~\cite{bitergia2016gender, gerosa2021shifting, trinkenreich2023belong}. For our analysis, we grouped those who self-identified as women or non-binary under a single category due to the small sample of non-binary participants (0.8\%). We assigned a code of 0 for males and 1 for minority groups.

\textbf{Control variables:} For the variable \textit{OSSCourse}, we asked participants a simple yes/no question: \textit{``Have you ever taken a course about OSS?''}. Regarding \textit{Involvement with OSS}, we categorized participants based on their responses to different questions. First, we asked, \textit{``Have you ever contributed to an OSS project?''}, participants who answered no were classified as \textit{never contributed}. For those who answered yes, we followed up with the question, \textit{``Do you still contribute to OSS projects?''}. Those who responded no were classified as \textit{dropped out}, while those who responded yes were classified as \textit{contributors}.

\vspace{-2px}
\subsection{Survey Design}
\vspace{-2px}

The survey instrument was structured in three parts, described in the following:

%\textit{\textbf{Population and inclusion criteria.}} Our study's target population is students in computer science and related fields.

%\textit{\textbf{Survey instrument.}} 
%The survey consisted of three main parts: (i) Part I -- Free and Open Source projects: concepts and perceptions; (ii) Part II -- Contributing to an OSS project; (iii) Part III -- Demographics. 
\textbf{Part I (perceptions):} 
In the first part of the survey, we collected student perceptions of barriers and motivational factors. They were also asked how they perceive OSS, with items that include motivations to join OSS projects and barriers reported in previous studies~\cite{gerosa2021shifting, balali2018newcomers, steinmacher2015social}. Table~\ref{tab:surveyquestions} describes the constructs and the questions used in the survey. As mentioned above, we used a 5-point Likert scale item for each indicator. % with five-point. %(ranging from 1 -- strongly disagree to 5 -- strongly agree).
The order of the items in each multi-item scale was randomized to mitigate primacy and recency effects~\cite{ralph2005pandemic}. Finally, participants were asked whether they were interested in contributing to OSS projects with a ten-point scale (ranging from 0---not interested---to 10---very interested).

\textbf{Part II (contributing):} Participants were asked whether they had already contributed to OSS projects. This helped us classify respondents into three groups: (i) students who had never contributed to OSS, (ii) students who had contributed to OSS but were no longer active, and (iii) students who were active OSS contributors. Additionally, we included an open question asking what OSS projects should do to make contributing more attractive to them.

\textbf{Part III (demographics):} Participants were asked to provide information about their gender, age, major, year of study, country of residence, and whether they have taken any courses about OSS.

Besides these three main sections, the survey includes a consent page that outlines the purpose of the research, the confidentiality rules, the estimated time required to complete the survey, and the researcher's contact information. We have also incorporated pre-screening validation questions to confirm that the participants belong to the target population. For this purpose, we adapted questions from Danilova et al.~\cite{danilova2021you}. Still, we included an attention check question to ensure participants read the instructions carefully. Although the item order was randomized, we did not change the order of the question blocks to ensure a clear flow for all respondents. Furthermore, we did not use URL tracking or collect any contact information, to maintain participant anonymity. Our survey instrument was provided in English and is included in the replication package~\cite{replicationpackage}.

\vspace{-2px}
\subsection{Data Collection and Analysis}
\vspace{-2px}

\textit{\textbf{Recruitment.}} We focused on increasing the sample's size and diversity by recruiting participants using the prolific platform~\cite{prolific}. Prolific is a crowdsourcing platform and has more than 150,000 active users~\cite{russo2022recruiting}. We conducted our survey using Prolific, considering the steps followed by some works that successfully recruited participants using this platform~\cite{russo2022recruiting, danilova2021you}. Prolific provides some pre-screening questions to help to narrow down the relevant populations. As our target population is computer science students, we selected the following filters: (i) student status; (ii) education level, filtered for undergraduate students; (iii) subject, filtered for computer science and computing (IT); and (iv) programming skills. At the time of the study, according to the Prolific information, we had 1,316 matching participants active on the platform for the past 90 days. We included questions in our survey to confirm that participants were part of our target population. Participants were paid \$3.0 (USD) through Prolific after completing the survey. 

\textit{\textbf{Filtering.}} We carefully reviewed and filtered our data to include only valid responses. All responses passed the attention check question, and none had the same choice for all Likert scale questions. We removed participants who did not complete the entire survey (13 cases), who failed the prescreening questions (50 cases), and who indicated they were no longer students (12 cases). Furthermore, we analyzed the time to complete the survey to remove lower outliers (0 removed). After filtering, we had 241 valid responses for analysis.
%The survey platform (Qualtrics~\cite{qualtrics}) had mechanisms to prevent multiple responses from the same participant.

We conducted a power analysis using the G*Power tool~\cite{faul2009statistical} to establish an appropriate sample size. Our model's maximum number of predictors is six (four latent variables and two control variables). The calculation indicated a minimum sample size of 98---our sample of 241 exceeded that number considerably. The effect size was moderate $f^2 = 0.15$, based on the PLS-SEM guidelines. We specified PLS-SEM as the statistical test due to its suitability for complex models.

%We conducted a power analysis using \textit{G*Power} to assess the statistical validity of our PLS-SEM analysis. Sensitivity analysis revealed that the minimum detectable effect size ($f^2$) for a sample size of 241 and a power of 0.80 is 0.058. Post-hoc analysis showed an achieved power of 0.998 for an effect size of $f^2 = 0.15$, confirming that the study is highly powered. Additionally, an a priori analysis indicated that a sample size of 98 would be sufficient to detect a medium effect size ($f^2 = 0.15$) with a power of 0.80. Thus, our sample size of 241 exceeds the required threshold, ensuring robust statistical validity.

\textit{\textbf{Analysis.}} We used the SmartPLS (v4.1.0) software~\cite{smartpls2024} for our analyses. The analysis procedures for PLS-SEM consist of two main steps, each involving specific tests and procedures. The first step involves evaluating the measurement model, which empirically assesses the relationships between the constructs and indicators (see Section~\ref{sec:evalmeasurementmodel}). The second step consists of evaluating the theoretical (or structural) model, which represents the hypotheses (see Section~\ref{sec:evaltheoreticalmodel}).

Furthermore, we qualitatively analyzed participants' comments on the open question about what OSS projects should do to make contributing to OSS more attractive. We employed a rigorous qualitative coding process, following open coding procedures~\cite{strauss1998basics}. The process was conducted using continuous comparison and discussion until reaching a consensus. Two researchers jointly analyzed the sets of answers to establish common ground, discussing the applied codes and disagreements until reaching a consensus. Finally, a third researcher discussed the classification until reaching the final outcome.

%\section{Participant Demographics}
%As seen in Table~\ref{tab:demographics},

\textit{\textbf{Demographics.}} Table~\ref{tab:demographics} shows that most participants had never contributed to OSS projects (70.5\%), 10.8\% contributed but dropped out, and 18.6\% were current contributors. Concerning gender distribution, most respondents identified as men (75.9\%), followed by women (22.4\%), which is consistent with the general distribution of students in the field. Regarding age, as also expected, a significant portion of respondents were 24 years old or younger (68.4\%). The largest group was computer science students (44.8\%), among other computing-related fields. Respondents were fairly distributed across their years of study, with juniors (32.8\%) and seniors (29\%) making up the majority of the sample. Furthermore, we received answers from diverse regions, but mainly from Europe (67.6\%).

%\begin{comment}
\begin{table}[!ht]\scriptsize
\centering
\caption{Respondent demographics (N=241).}
\label{tab:demographics}
\vspace{-4px}
\begin{tabular}{lr|lr}
\hline
\multicolumn{2}{c|}{\textbf{Profile}} & \multicolumn{2}{c}{\textbf{Major}} \\ \hline
\multicolumn{1}{l|}{Never contributed} & \textbf{70.5\%} & \multicolumn{1}{l|}{Applied computer science} & \textbf{5\%} \\ \hline
\multicolumn{1}{l|}{Dropped out} & \textbf{10.8\%} & \multicolumn{1}{l|}{Computer engineering} & \textbf{9.9\%} \\ \hline
\multicolumn{1}{l|}{Contributor} & \textbf{18.6\%} & \multicolumn{1}{l|}{Computer science} & \textbf{44.8\%} \\ \hline
\multicolumn{2}{c|}{\textbf{Gender}} & \multicolumn{1}{l|}{Information systems} & \textbf{5.4\%} \\ \hline
\multicolumn{1}{l|}{Man} & \textbf{75.9\%} & \multicolumn{1}{l|}{Information technology} & \textbf{13.2\%} \\ \hline
\multicolumn{1}{l|}{Woman} & \textbf{22.4\%} & \multicolumn{1}{l|}{Software engineering} & \textbf{11.2\%} \\ \hline
\multicolumn{1}{l|}{Non-binary} & \textbf{0.8\%} & \multicolumn{1}{l|}{\begin{tabular}[c]{@{}l@{}}Other computing-related \\ major/program\end{tabular}} & \textbf{10.3\%} \\ \hline
\multicolumn{1}{l|}{Prefer not to say} & \textbf{0.8\%} & \multicolumn{2}{c}{\textbf{Year of Study}} \\ \hline
\multicolumn{2}{c|}{\textbf{Age}} & \multicolumn{1}{l|}{1st year (freshman)} & \textbf{11.6\%} \\ \hline
\multicolumn{1}{l|}{24 or below} & \textbf{68.4\%} & \multicolumn{1}{l|}{2nd year (sophomore)} & \textbf{26.5\%} \\ \hline
\multicolumn{1}{l|}{25 to 34} & \textbf{18.6\%} & \multicolumn{1}{l|}{3rd year (junior)} & \textbf{32.8\%} \\ \hline
\multicolumn{1}{l|}{35 to 44} & \textbf{2.5\%} & \multicolumn{1}{l|}{4th or 5th year (senior)} & \textbf{29\%} \\ \hline
\multicolumn{1}{l|}{45 to 54} & \textbf{0.4\%} & \multicolumn{2}{c}{\textbf{Continent of Residence}} \\ \hline
\multicolumn{1}{l|}{Prefer not to say} & \textbf{9.9\%} & \multicolumn{1}{l|}{Africa} & \textbf{16.5\%} \\  \hline\cline{1-2}
 & \multicolumn{1}{l|}{\textbf{}} & \multicolumn{1}{l|}{Asia} & \textbf{1.2\%} \\ \cline{3-4} 
 & \multicolumn{1}{l|}{\textbf{}} & \multicolumn{1}{l|}{Europe} & \textbf{67.6\%} \\ \cline{3-4} 
 & \multicolumn{1}{l|}{\textbf{}} & \multicolumn{1}{l|}{North and Central America} & \textbf{13.2\%} \\ \cline{3-4} 
 & \multicolumn{1}{l|}{\textbf{}} & \multicolumn{1}{l|}{South America} & \textbf{1.2\%} \\ \cline{3-4} \cline{3-4} 
 %\hline \hline
\end{tabular}
\vspace{-5px}
\end{table}
%\end{comment}

\textit{\textbf{Replication package.}} Our replication package provides the anonymized dataset, instruments, and scripts~\cite{replicationpackage}.

\begin{figure*}[!ht]
    \centering
    \includegraphics[width=15cm]{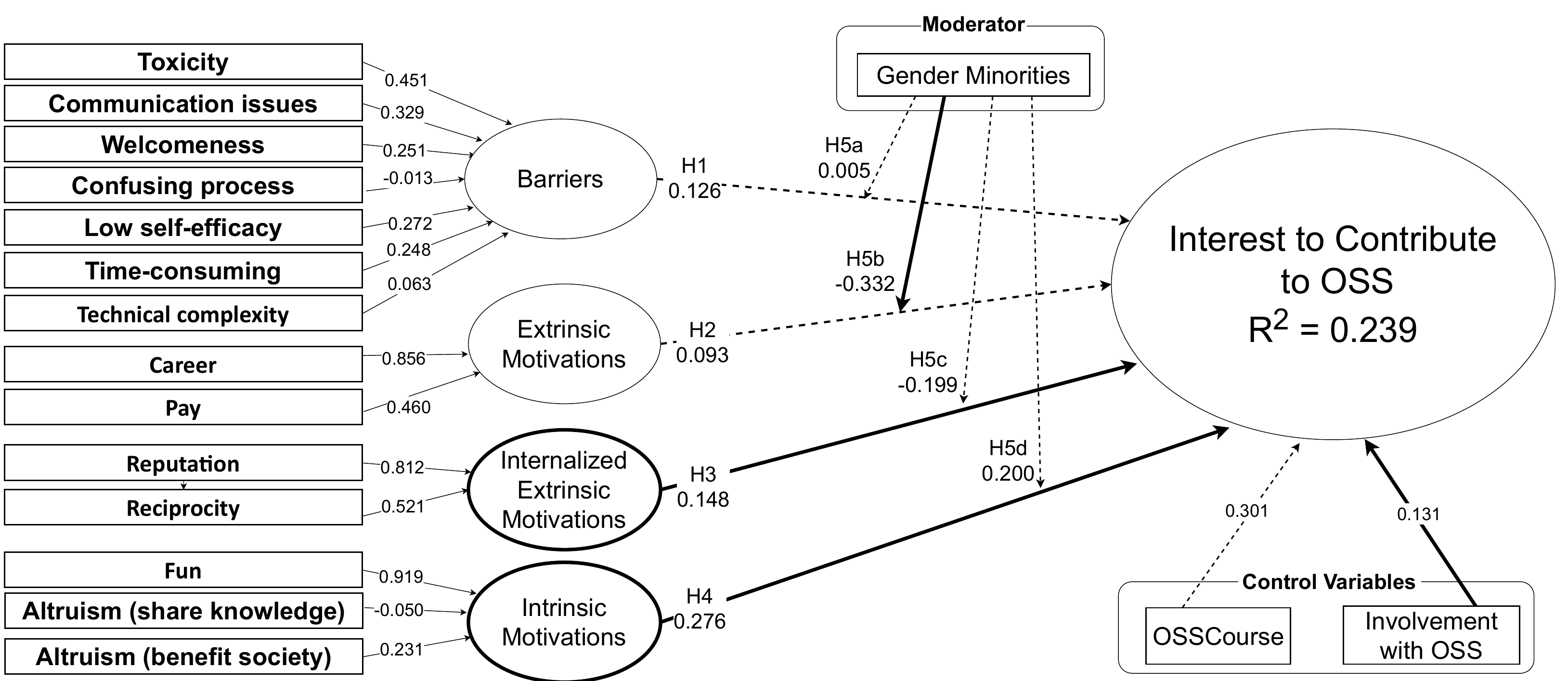}
    \caption{Item loadings and path coefficients at \(p < 0.05\) are shown with full lines; non-significant links with dashed lines.}
    \label{fig:modelpls}
    \vspace{-10px}
\end{figure*}

\vspace{-5px}
\section{Analysis and Results}
\label{sec:results}
\vspace{-2px}

%\vspace{-2px}
\subsection{Factors associated with students' interest (RQ1)}
\label{sec:rq1}
\vspace{-2px}

To identify factors associated with students' interest in contributing to OSS, we analyzed our theoretical model, beginning with the evaluation of the measurement model (Section~\ref{sec:evalmeasurementmodel}), followed by the hypotheses evaluation within the structural model (Section~\ref{sec:evaltheoreticalmodel}). We assessed the significance of our model using the protocol proposed by previous research~\cite{russo2021pls, hair2019use}. The path weighting scheme was estimated using SmartPLS 4~\cite{sarstedt2019partial}. 

Our model comprises four exogenous variables: barriers, extrinsic, internalized extrinsic, and intrinsic motivation. Additionally, the model incorporates the moderators \textit{Gender} and the control variables \textit{OSSCourse} and \textit{Involvement with OSS}. We hypothesized that these exogenous variables are associated with the endogenous variable \textit{Interest in contributing to OSS}. 
%We evaluated the hypotheses using PLS-SEM. \\

\subsubsection{\textbf{Evaluation of the Measurement Model}}
\label{sec:evalmeasurementmodel}

Some of the constructs in the theoretical model (see Figure~\ref{fig:modelpls}) are modeled as latent variables, meaning they are measured by multiple indicators (items/questions on the survey). The first step in evaluating a structural equation model is to assess the soundness of the measurement of these latent variables---this process is known as evaluating the ``measurement model''~\cite{hair2019use}. 
%We present the assessment of several criteria in this evaluation.
The constructs \textit{Barriers, Extrinsic, Internalized Extrinsic, and Intrinsic} comprise our formative measurement model. We validated our model through the tests described below:
%According to Russo and Stol~\cite{russo2021pls}, the formative model is validated through tests that include content validity, convergent validity, collinearity, and indicator weights’ significant reliability.

%\textbf{Content validity.} Russo and Stol~\cite{russo2021pls} emphasize the importance of meticulous content specification, where the domain content of the measured items is clearly defined, as these items are neither replaceable nor interchangeable~\cite{russo2021pls}. To promote content validity, our survey employed instruments and constructs from previous literature that focused on barriers~\cite{steinmacher2015social} and motivations~\cite{von2012carrots}.

%Content validity is important when dealing with formative measurement models. Russo and Stol~\cite{russo2021pls} suggest that the highest care should be devoted to the content specification in which it is specified the domain’s content of the measured items since they are not replaceable and interchangeable~\cite{russo2021pls}. To guarantee content validity, our survey used validated scales in previous literature reviews that focused on barriers~\cite{steinmacher2015social} and motivations~\cite{von2012carrots}.

\textbf{Indicators collinearity.} Unlike reflective indicators, which are essentially interchangeable, high correlations are not expected between items in formative measurement models. High correlations between two formative indicators demonstrate collinearity. The variance inflation factor (VIF) should be less than 5~\cite{hair2021primer}. In our model, all indicators had a VIF lower than the threshold (see Table~\ref{tab:outerweights}).

%The construct indicators were considered valid, as their outer weights were significant (management innovation), or their outer loadings were above 0.50 and were significant (service, process, and marketing innovation) (Hair et al., 2017). Although the outer weight of the product innovation indicator was insignificant and the outer loading was relatively low (0.376), the item was retained, as the outer loading was significant and the indicator is theoretically relevant to the innovation construct (Hair et al., 2017).

\textbf{Indicator weights’ significant reliability.} According to Hair et al.~\cite{hair2021primer}, the bootstrapping analysis relies on the weight of indicators (relative importance of the indicator) and the analysis of their loadings (absolute importance of the indicator) to assess the significance of each indicator. %At a five percent significance level, bootstrapping results should be at least 1.96 to justify retaining a formative indicator in the model. 
When an indicator's outer weight is nonsignificant, but its outer loading is relatively high (above 0.50), it is generally advisable to retain the indicator. However, if an indicator has a nonsignificant weight and an outer loading below 0.50, researchers should evaluate the p-value of the outer loading to determine whether the indicator should be retained or removed~\cite{hair2021primer}. %If both weight and loading values are insignificant, the indicator is deemed irrelevant and can be removed from the model. 
Using the recommendations of Hair et al.~\cite{hair2021primer}, we followed an iterative process to evaluate the indicators of our constructs. We removed indicators when their measurements were both insignificant, as this suggested that the indicators were not relevant to the construct. Specifically, two indicators from the construct \textit{internalized extrinsic} (Own-Use and Learning) and two indicators from the construct \textit{intrinsic} (Kinship and Ideology) were removed.
% Although some indicators had low weight values, they were retained in the adjusted model due to their significant loading values. 
%The relevant indicators remaining in the model represent significant constructs based on the data collected from the sample population. 
Table~\ref{tab:outerweights} summarizes the results for the formative measured constructs. Note that the removed indicators are not shown due to the iterative process since, with each indicator removal, the values are recalculated. 

\begin{table*}[!ht]\scriptsize
\centering
\caption{Formative constructs measurement model testing results.}
\label{tab:outerweights}
\vspace{-4px}
\begin{tabular}{l|l|c|c|c|c|c}
\hline
\multicolumn{1}{c|}{\textbf{Constructs}} & \multicolumn{1}{c|}{\textbf{Indicators}} & \textbf{VIF} & \textbf{\begin{tabular}[c]{@{}c@{}}Outer weights \\ {[}Outer loadings{]}\end{tabular}} & \textbf{\(t\) Value} & \textbf{\begin{tabular}[c]{@{}c@{}}\(p\) Value Outer weights \\ {[}Outer loadings{]}\end{tabular}} & \textbf{\begin{tabular}[c]{@{}c@{}}Confidence interval \\ 95\% percentile\end{tabular}} \\ \hline \hline

\multirow{7}{*}{Barriers} & Toxicity & 1.167 & 0.451 {[}0.714{]} & 2.456 & 0.014 [0] & {[}0.026, 0.739{]} \\ \cline{2-7} 

 & Communication issues & 1.227 & 0.329 {[}0.634{]} & 1.668 & 0.095 [0] & {[}-0.117, 0.664{]} \\ \cline{2-7} 
 
 & Welcomeness & 1.203 & 0.251 {[}0.581{]} & 1.276 & 0.202 [0] & {[}-0.216, 0.549{]} \\ \cline{2-7} 
 
 & Confusing process & 1.443 & -0.013 {[}0.485{]} & 0.053 & 0.958 [0.009] & {[}-0.526, 0.451{]} \\ \cline{2-7} 
 
 & Low self-efficacy & 1.219 & 0.272 {[}0.574{]} & 1.241 & 0.215 [0.001] & {[}-0.188, 0.666{]} \\ \cline{2-7} 
 
 & Time-consuming & 1.375 & 0.248 {[}0.583{]} & 1.053 & 0.292 [0] & {[}-0.257, 0.652{]} \\ \cline{2-7} 
 
 & Technical complexity & 1.371 & 0.063 {[}0.467{]} & 0.251 & 0.802 [0.01] & {[}-0.423, 0.563{]} \\ \hline \hline
 
\multirow{2}{*}{Extrinsic motivation} & Career & 1.005 & 0.856 {[}0.889{]} & 2.956 & 0.003 [0.001] & {[}-0.09, 1.004{]} \\ \cline{2-7} 

 & Pay & 1.005 & 0.46 {[}0.521{]} & 1.224 & 0.221 [0.153] & {[}-0.444, 0.988{]} \\ \hline \hline
 
\multirow{2}{*}{Internalized extrinsic motivation} & Reputation & 1.007 & 0.812 {[}0.855{]} & 5.223 & 0 [0] & {[}0.425, 0.999{]} \\ \cline{2-7} 

 & Reciprocity & 1.007 & 0.521 {[}0.587{]} & 2.25 & 0.024 [0.01] & {[}-0.044, 0.869{]} \\ \hline \hline
 
\multirow{3}{*}{Intrinsic motivation} & Fun & 1.207 & 0.919 {[}0.978{]} & 10.469 & 0 [0] & {[}0.698, 1.037{]} \\ \cline{2-7} 

 & Altruism (share knowledge) & 1.289 & -0.05 {[}0.375{]} & 0.315 & 0.753 [0.006] & {[}-0.349, 0.283{]} \\ \cline{2-7} 
 
 & Altruism (benefit society) & 1.265 & 0.231 {[}0.518{]} & 1.487 & 0.137 [0] & {[}-0.074, 0.53{]} \\ \hline \hline
\end{tabular}
\vspace{-10px}
\end{table*}

\subsubsection{\textbf{Evaluation of the Theoretical Model}}
\label{sec:evaltheoreticalmodel}

We now discuss the evaluation of the theoretical model.

\textbf{\textit{Assessing Collinearity.}} Our theoretical model includes four exogenous variables: barriers and extrinsic, internalized extrinsic, and intrinsic motivations. Additionally, the model incorporates the moderators \textit{Gender} and the control variables \textit{OSSCourse} and \textit{Involvement with OSS}. We hypothesize that these exogenous variables are associated with the endogenous variable \textit{Interest in contributing to OSS}. To ensure the independence of the four exogenous constructs, we calculated their collinearity using VIF. To avoid collinearity problems, VIF values should be under 5.0~\cite{hair2021primer}. In the adjusted model, all constructs have VIF values below this threshold, %as shown in Table~\ref{tab:vif}, 
confirming that collinearity is not a concern in our model. 
%, allowing for a reliable assessment of the relationships between the exogenous and endogenous variables.

\begin{comment}
\begin{table}[!ht]\scriptsize
\centering
\caption{VIF values for exogenous constructs.}
\label{tab:vif}
\vspace{-4px}
\begin{tabular}{l|l}
\hline
\textbf{Construct} & {\color[HTML]{000000} \textbf{VIF}} \\ \hline \hline

Barriers & {\color[HTML]{000000} 1.681} \\ \hline
Extrinsic & {\color[HTML]{000000} 1.454} \\ \hline
Gender & {\color[HTML]{000000} 1.101} \\ \hline
Internalized Extrinsic & {\color[HTML]{000000} 1.592} \\ \hline
Intrinsic & {\color[HTML]{000000} 1.843} \\ \hline
OSSCourse & {\color[HTML]{000000} 1.083} \\ \hline
ProfileType & {\color[HTML]{000000} 1.112} \\ \hline
Gender x Extrinsic & {\color[HTML]{000000} 1.482} \\ \hline
Gender x Barriers & {\color[HTML]{000000} 1.865} \\ \hline
Gender x Intrinsic & {\color[HTML]{000000} 1.957} \\ \hline
Gender x Internalized Extrinsic & {\color[HTML]{000000} 1.577} \\ \hline \hline
\end{tabular}
\vspace{-10px}
\end{table}
\end{comment}

\textbf{\textit{Path coefficients and significance.}} %According to Hair et al.~\cite{hair2021primer}, PLS does not make strong assumptions about the data distribution, so parametric significance tests should not be used. Instead, 
PLS employs a bootstrapping procedure to evaluate the significance of path coefficients. This involves drawing a large number (typically five thousand) of random subsamples with replacement~\cite{trinkenreich2023belong}. Replacement ensures that all subsamples have the same number of observations as the original dataset, and the path model is estimated.% for each subsample.

A standard error can be determined from the resulting bootstrap distribution, which is used to make statistical inferences~\cite{hair2021primer}. The mean path coefficient determined through bootstrapping may differ slightly from the path coefficient calculated directly from the sample. The variability is captured in the standard error of the sampling distribution of the mean. This approach allows for robust statistical evaluation without relying on parametric assumptions. Table~\ref{tab:pathcoef} presents the results for our hypothesis evaluations. 
%, including the path coefficients, standard deviations (SD), 95\% confidence intervals, t-values, and p-values.

The path coefficients in Figure~\ref{fig:modelpls} and Table~\ref{tab:pathcoef} are interpreted as standardized regression coefficients, indicating the direct effect of a variable on another~\cite{hair2021primer}. Each hypothesis is represented
by an arrow in the diagram in Figure~\ref{fig:modelpls}. For example, the arrow pointing from Intrinsic motivation to \textit{Interest to contribute to OSS} represents H4. A positive path coefficient of 0.276 suggests that intrinsic motivation is positively associated with an interest in contributing to OSS, meaning when \textit{Intrinsic} motivation increases by one standard deviation unit, \textit{Interest to contribute to OSS} increases by 0.276 standard deviation unit. The standard deviation represents the amount of variation within a set of values.

Based on these results, we found support for hypotheses H3 (\(p = 0.041\)), H4 (\(p = 0.001\)), and H5b (\(p = 0.047\)). Among the two control variables, only \textit{Involvement with OSS} showed a significant association with \textit{Interest to contribute to OSS} (\(p = 0.007\)).

\begin{table*}[!hbt]\scriptsize
\centering
\caption{Standardized path coefficients, standard deviations, confidence intervals, and p values.}
\label{tab:pathcoef}
\vspace{-4px}
\begin{tabular}{l|c|c|c|c|c|c}
\cline{2-7}

\multicolumn{1}{c}{\textbf{}} & \textbf{\begin{tabular}[c]{@{}c@{}}Path \\ Coefficients\end{tabular}} & \textbf{SD} & \textbf{\begin{tabular}[c]{@{}c@{}}Confidence \\ Interval 95\%\end{tabular}} & \textbf{\(t\) Value} & \textbf{\(p\) Value} & \textbf{\begin{tabular}[c]{@{}c@{}}Significance \\ (\(p < 0.05\))?\end{tabular}} \\ \hline \hline

\multicolumn{1}{l|}{H1: Barriers $\rightarrow$ Interest in OSS} & -0.126 & 0.08 & (-0.232, 0.066) & 1.577 & 0.115 & No \\ \hline

\multicolumn{1}{l|}{H2: Extrinsic $\rightarrow$   Interest in OSS} & 0.093 & 0.069 & (-0.06, 0.215) & 1.34 & 0.18 & No \\ \hline

\multicolumn{1}{l|}{\cellcolor{green!75} 
 \textbf{H3: Internalized Extrinsic $\rightarrow$ Interest in OSS}} & \cellcolor{green!75} \textbf{0.148} & \cellcolor{green!75} \textbf{0.072} & \cellcolor{green!75} \textbf{(0.008, 0.287)} & \cellcolor{green!75} \textbf{2.043} & \cellcolor{green!75} \textbf{0.041} & \cellcolor{green!75} \textbf{Yes} \\ \hline

\multicolumn{1}{l|}{\cellcolor{green!75} 
 \textbf{H4: Intrinsic $\rightarrow$ Interest in OSS}} & \cellcolor{green!75} \textbf{0.276} & \cellcolor{green!75} \textbf{0.079} & \cellcolor{green!75} \textbf{(0.126, 0.438)} & \cellcolor{green!75} \textbf{3.472} & \cellcolor{green!75} \textbf{0.001} & \cellcolor{green!75} \textbf{Yes} \\ \hline \hline

\multicolumn{1}{l|}{H5a: Gender minorities $\times$ Barriers $\rightarrow$  Interest in OSS} & 0.005 & 0.18 & (-0.328, 0.379) & 0.03 & 0.976 & No \\ \hline

\multicolumn{1}{l|}{\cellcolor{green!75} 
 \textbf{H5b: Gender minorities$\times$ Extrinsic $\rightarrow$   Interest in OSS}} & \cellcolor{green!75} \textbf{-0.332} & \cellcolor{green!75} \textbf{0.167} & \cellcolor{green!75} \textbf{(-0.689, -0.056)} & \cellcolor{green!75} \textbf{1.991} & \cellcolor{green!75} \textbf{0.047} & \cellcolor{green!75} \textbf{Yes} \\ \hline

\multicolumn{1}{l|}{H5c: Gender minorities $\times$ Internalized Extrinsic $\rightarrow$ Interest in OSS} & -0.199 & 0.174 & (-0.486, 0.189) & 1.146 & 0.252 & No \\ \hline
\multicolumn{1}{l|}{H5d: Gender minorities $\times$ Intrinsic $\rightarrow$   Interest in OSS} & 0.2 & 0.17 & (-0.128, 0.53) & 1.181 & 0.238 & No \\ \hline \hline

\multicolumn{1}{l|}{OSSCourse $\rightarrow$ Interest in OSS} & 0.301 & 0.186 & (-0.066, 0.66) & 1.616 & 0.106 & No \\ \hline

\multicolumn{1}{l|}{\cellcolor{green!75} 
 \textbf{Involvement with OSS $\rightarrow$ Interest in OSS}} & \cellcolor{green!75} \textbf{0.131} & \cellcolor{green!75} \textbf{0.048} & \cellcolor{green!75} \textbf{(0.042, 0.229)} & \cellcolor{green!75} \textbf{2.722} & \cellcolor{green!75} \textbf{0.007} & \cellcolor{green!75} \textbf{Yes} \\ \hline \hline
\end{tabular}
\vspace{-10px}
\end{table*}

\textbf{\textit{Coefficient of determination.}} We assessed the relationships between constructs and the model's predictive capabilities. The coefficient of determination ($R^2$ value) is the most common measure for evaluating structural models and defines the predictive accuracy of the model~\cite{hair2021primer}. In our model, the $R^2$ value for the endogenous variable (\textit{Interest to contribute to OSS}) was 0.239, as illustrated in Figure~\ref{fig:modelpls}. According to Russo and Stol~\cite{russo2021pls}, the software engineering community has not yet reached a consensus on thresholds for $R^2$ values as the value should be interpreted in light of the model complexity and research discipline. Using the general guidelines by Hair et al.~\cite{hair2021primer} and Henseler et al.~\cite{henseler2009use}, this value is considered weak-moderate. 
%Thus, while our $R^2$ value suggests a relatively low level of predictive accuracy, its interpretation should be contextualized within the broader discussion on model evaluation standards in software engineering research.

We also inspected Stone-Geisser’s $Q^2$ value~\cite{stone1974cross}, which measures external validity and indicates the model’s predictive relevance~\cite{hair2021primer}. The $Q^2$ value can be obtained through a blindfolding procedure available in the SmartPLS software. Blindfolding is a resampling technique that omits certain data points, predicts the omitted values, and then uses the prediction error to cross-validate the model estimates~\cite{tenenhaus2005pls}. In our model, $Q^2$ values were calculated for \textit{Interest to contribute to OSS}, the reflective endogenous construct, yielding a value of 0.16. According to the guidelines~\cite{hair2019use}, $Q^2$ values greater than 0 indicate the construct has predictive relevance. In contrast, negative values suggest that the model does not perform better than the simple average of the endogenous variable. A $Q^2$ value of 0.16 suggests that our model has predictive relevance.

The Standardized Root Mean Square Residual (SRMR) is a widely used fit measure appropriate for detecting misspecification in PLS-SEM models~\cite{russo2021pls}. SRMR represents the square root of the sum of the squared differences between the model-implied and empirical correlation matrices, essentially measuring the Euclidean distance between these two matrices~\cite{henseler2014common}. An SRMR value of 0 indicates a perfect model fit. In our analysis, the SRMR was 0.06, which falls within the acceptable fit threshold of 0.08, as recommended by Henseler et al.~\cite{henseler2016using}. Therefore, our model demonstrates a good fit, supporting the validity of the theoretical model.

\textbf{\textit{Moderating factors.}} We examined our data to determine if the impact of barriers, extrinsic motivations, internalized extrinsic motivations, and intrinsic motivations on the interest in contributing to OSS varies by gender (i.e., men and minorities). We report only the significant results at 0.05, with confidence intervals calculated through bootstrapping.

The \textbf{gender} variable significantly moderates by reducing the association between extrinsic motivation and interest to contribute to OSS when being part of gender minorities, supporting the H5b hypothesis. However, we did not find significant moderation of \textit{gender} on the association between the other constructs (\textit{Barriers, Internalized extrinsic motivations, Intrinsic motivations}) and Interest in OSS. Therefore, our results do not support H5a, H5c, and H5d hypotheses.

%Figure~\ref{fig:slope} illustrates the interaction effect between gender and extrinsic motivation on Interest in OSS (InterestOSS). The two lines represent different levels of the gender variable: the Red Line -- Gender at zero -- represents the gender male, and the Green Line -- Gender at one --represents the minority group. The red line has a positive slope, indicating that for the male group, an increase in extrinsic motivation is associated with a slight increase in InterestOSS. Meanwhile, the green line has a negative slope, indicating that for the minority group represented an increase in extrinsic motivation is associated with a decrease in InterestOSS. Moreover, the intersection of the two lines suggests that the effect of extrinsic motivation on InterestOSS differs by gender.

\begin{comment}  
\begin{figure}[ht]
    \centering
    \includegraphics[width=8.5cm]{images/slopegraph.png}
    \caption{Gender as a moderator of Extrinsic motivation $\rightarrow$ InterestOSS.}
    \label{fig:slope}
\end{figure}
\end{comment}

\begin{table*}[]\scriptsize
\centering
\caption{Students' recommendations to make OSS projects more attractive.}
\label{tab:studentsrecomendossatractive}
\vspace{-4px}
\begin{tabular}{l|l|c|c|c}
\hline
\multicolumn{1}{c|}{\textbf{Category}} & \multicolumn{1}{c|}{\textbf{Themes}} & \textbf{\begin{tabular}[c]{@{}c@{}}Never contributed (N=170)\end{tabular}} & \textbf{\begin{tabular}[c]{@{}c@{}}Dropped out (N=26)\end{tabular}} & \textbf{\begin{tabular}[c]{@{}c@{}}Contributor (N=45)\end{tabular}} \\ \hline \hline

\multirow{7}{*}{\begin{tabular}[c]{@{}l@{}}Contribution Process \\ Improvement (7)\end{tabular}} & Simplify the contribution process & \textbf{32 (18.8\%)} & \textbf{5 (19.2\%)} & \textbf{8 (17.8\%)} \\ \cline{2-5} 
 & Well-documented OSS project & \textbf{12 (7.1\%)} & \textbf{2 (7.7\%)} & \textbf{2 (4.4\%)} \\ \cline{2-5} 
 & Mentoring & \textbf{10 (5.9\%)} & \textbf{1 (3.8\%)} & \textbf{1 (2.2\%)} \\ \cline{2-5} 
 & Label tasks & \textbf{7 (4.1\%)} & \textbf{-} & \textbf{2 (4.4\%)} \\ \cline{2-5} 
 & User-friendly interface & \textbf{2 (1.2\%)} & \textbf{2 (7.7\%)} & \textbf{4 (8.9\%)} \\ \cline{2-5} 
 & Gamify the contribution process & \textbf{-} & \textbf{-} & \textbf{3 (6.7\%)} \\ \cline{2-5} 
 & Provide a channel to ask questions & \textbf{3 (1.8\%)} & \textbf{-} & \textbf{1 (2.2\%)} \\ \hline \hline
 
\multirow{4}{*}{\begin{tabular}[c]{@{}l@{}}Awareness and \\ Outreach (4)\end{tabular}} & Encourage students & \textbf{8 (4.7\%)} & \textbf{8 (30.8\%)} & \textbf{6 (13.3\%)} \\ \cline{2-5} 
 & OSS Attractiveness is not a problem & \textbf{15 (8.8\%)} & \textbf{-} & \textbf{4 (8.9\%)} \\ \cline{2-5} 
 & Advertise the OSS projects & \textbf{11 (6.5\%)} & \textbf{-} & \textbf{4 (8.9\%)} \\ \cline{2-5} 
 & Create welcome environment & \textbf{11 (6.5\%)} & \textbf{-} & \textbf{-} \\ \hline \hline
 
\multirow{3}{*}{Training (3)} & Create training projects & \textbf{4 (2.4\%)} & \textbf{-} & \textbf{1 (2.2\%)} \\ \cline{2-5} 
 & Provide a study environment & \textbf{4 (2.4\%)} & \textbf{1 (3.8\%)} & \textbf{-} \\ \cline{2-5} 
 & Introduce at university & \textbf{-} & \textbf{1 (3.8\%)} & \textbf{5 (11.1\%)} \\ \hline \hline
 
\multirow{3}{*}{\begin{tabular}[c]{@{}l@{}}Incentives and \\ Personalization (3)\end{tabular}} & Provide financial compensation & \textbf{12 (7.1\%)} & \textbf{2 (7.7\%)} & \textbf{-} \\ \cline{2-5} 
 & Reward the contributors & \textbf{9 (5.3\%)} & \textbf{1 (3.8\%)} & \textbf{2 (4.4\%)} \\ \cline{2-5} 
 & Aligned with personal interests & \textbf{4 (2.4\%)} & \textbf{-} & \textbf{-} \\ \hline \hline
 
\begin{tabular}[c]{@{}l@{}}Diversity and Inclusion (1)\end{tabular} & Be available in other languages & \textbf{1 (0.6\%)} & \textbf{-} & \textbf{-} \\ \hline \hline
\end{tabular}
\vspace{-10px}
\end{table*}

\textbf{\textit{Control Variables.}} We also examined our data to determine whether having a course or being a contributor affects the interest in contributing to OSS. As expected, previous experience affects interest. Being a contributor is associated with increased interest with a significantly positive effect. We did not find a significant association between having a course and interest.

\rqone[
    \tcblower
    \textbf{Answer:} The factors associated with students' interest in contributing to OSS projects include (i) intrinsic motivation (H4 supported); (ii) internalized extrinsic motivation (H3 supported); and (iii) gender moderation when perceiving extrinsic motivational factors (H5b supported). Additionally, students who are currently contributing to OSS have a higher interest in contributing than students who never contributed and dropped out.
]{}

\vspace{-2px}
\subsection{Students' recommendations (RQ2)}
\label{sec:rq2}
\vspace{-2px}

To identify recommendations students have to make projects more attractive, we adopted open coding in the survey responses. 
%This approach allowed us to identify common themes and understand students' opinions on making OSS projects more attractive. 
Table~\ref{tab:studentsrecomendossatractive} presents the results of the analysis, categorized by their involvement with OSS (i.e., never contributed, dropped out, and contributor). Each recommendation category lists the number of responses with their corresponding percentage of participants who suggested them.

\textbf{Contribution process improvement.} This category focuses on enhancing the processes and resources facilitating student contributions to OSS projects. The most frequently mentioned recommendation was to \textbf{simplify the contribution process}. The feedback from students highlights the importance of providing lightweight briefings and clear architecture, highlighted by P6's remark: ``\textit{Have the code simplified and easy for people with less skills to be able to help too.}'' %Moreover, easing the contribution process, as indicated by P218 ``\textit{They should be easily available to someone who does not know where to begin.}'' Another suggestion relates to offering detailed guides and beginner-friendly resources, giving clear instructions and readily available help, and reducing intimidation to make the process more accessible, as mentioned by P127, ``\textit{Make it less intimidating to contribute.}''
Another recommendation was the need for \textbf{well-documented OSS projects}. Key recommendations include providing comprehensive architecture summaries. %stated by P221 ``\textit{Perhaps some sort of starting guide on how to work within the OSS. Not everyone understands each other's code, so some sort of easy introduction into the project could be beneficial, see major companies and how they fire old staff and the new staff comes in and doesn't know how to code within that environment.}'' 
Moreover, students indicated the need for detailed guides, including ``getting started'' resources and lists of potential improvements. These were highlighted as essential for new contributors to understand where and how to begin. In addition, students indicated the need for properly structured README files and introductory documentation to help new contributors understand the project's architecture and development environment, easing their onboarding process. %as described by P8, ``\textit{They should be well documented, have proper READMEs and project structures, so that no matter the amount of experience somebody has with a certain project, or with contributing to open source in general, they can get started easily and understand the project's inner structure quickly.}''

Students also consider that \textbf{mentoring} plays a crucial role in attracting and retaining new contributors to OSS projects. Key recommendations include providing clear support, standards, and instructions and offering free training sessions focusing on junior developers, as outlined by P66: ``\textit{Having projects or groups that focus on helping junior developers learn more and develop their skills and guide them to how that specific project works would definitely attract me.}'' %In addition, providing support to reduce anxiety and help newcomers feel more comfortable can encourage participation, as stated in P83: ``\textit{I honestly don't know, I just have a lot of anxiety about working on something without knowing the other people working on it.}'' 
Furthermore, other suggestions include guiding newcomers on where to start, offering personalized tutoring, and combining improved documentation with mentorship.
Additionally, \textbf{labeling tasks} could help to provide the skills required for each task, helping newcomers assess whether they are ready to tackle specific issues. %as described in P82: ``\textit{They should allocate easier, beginner-level tasks to students and allow for patience as they navigate an unfamiliar environment. Overall, making the process more accepting would help students.}'' 
In addition, students mentioned that providing clear guidance on where to start would be important, allowing skill level selection and creating job/task boards as indicated by ``\textit{Put forward some kind of job/task board with skill level required.}'' Likewise, creating a \textbf{user-friendly interface} was also suggested by students that highlighted the importance of making the interface more user-friendly for beginners, providing quick explanations of interface elements. %, and offering comprehensive user guides. %, as outlined by P65: ``\textit{Being more user-friendly for beginners in OSS.}'' and P159: ``\textit{Quick and short explanation of a button when they hover their mouse.}''

\textbf{Gamifying the contribution process} was also suggested to enhance engagement and support for new contributors. Students recommend implementing levels based on contributors' knowledge and understanding and creating fun challenges to make the process more engaging, as highlighted by P40: ``\textit{Make levels i.e., allow students to contribute based on their level of knowledge and understanding.}'' Another recommendation was \textbf{providing a channel to ask questions} to facilitate the open dialogue among all participants to help create a supportive community where contributors feel comfortable asking questions and sharing information as mentioned by P170: ``\textit{Having a more open dialogue between all the participants.}''

\textbf{Awareness and outreach.} %This category emphasizes strategies to increase awareness and engagement with OSS projects. 
Recommendations related to this category include \textbf{encouraging students} through increasing the popularity and awareness of OSS projects, academic benefits, job opportunities, and social influence as highlighted by P20 ``\textit{I would need more friends doing them.}'' Moreover, we observed that for some students \textbf{OSS attractiveness is not a problem}, indicating that they find OSS projects inherently interesting and engaging. We also found students who think OSS is not for them, as suggested by P55 ``\textit{they are already attractive to me, it's just that I don't think I have enough skills to contribute.}'' and P172 ``\textit{I don't think they ``should'' do anything. OSS is mainly created \& used by professionals, students are not the target audience.}''

Effective \textbf{advertising of OSS projects} was also seen as a way to increase participation, with suggestions to make projects more engaging through fun advertising, 
%promoting them among end-users, reaching out to potential contributors, 
targeting lower skill levels, 
%frequent advertising, spreading more information, creating more projects, and 
highlighting the benefits of contributing on resumes and professional platforms as indicated by P147: ``\textit{Promote it on sites like LinkedIn, so students can add the information about the contribution to their profiles.}'' % and outlined by P91: ``\textit{I think OSS projects could describe how good they could look on a resume and how companies could perceive it as good experience.}''.
Creating a \textbf{welcoming environment} by hosting events and being more inclusive to new and inexperienced members was also recommended. %, as mentioned by P131: ``\textit{Have a more welcoming community or have some more learning-based tasks for newcomers as OSS communities do not look as beginner-friendly due to the scope.}'' 
%Organizing open meetings where the community can talk and work together can enhance collaboration and a sense of community and foster a positive atmosphere. %, as described in P174: ``\textit{PEOPLE, atmosphere around it, not being rude when somebody doesn't know how to start, when somebody has lower skills than others.}''

\textbf{Training.} 
%This category highlights the need for training and educational initiatives to support students in contributing to OSS projects. 
Students recommended that \textbf{creating training projects} could be used as a strategy to attract and support new contributors to OSS projects. Key recommendations include organizing beginner training sessions, developing beginner-friendly projects with detailed guides, grouping projects by difficulty, providing mentorship, and designing skill development projects as mentioned by P22: ``\textit{Create projects that students could join and learn more and increase their skills.}'' In addition, creating a supportive \textbf{study environment} appeared as crucial for encouraging students to contribute to OSS projects. Key recommendations include developing beginner-friendly tutorials and documentation. %, as outlined by P42: ``\textit{Create beginner-friendly YouTube tutorials and documentation that allow the potential contributor to have a good feel of what they would like to work on, as far as a particular project is concerned.}.'' 
Furthermore, \textbf{introducing OSS projects within the university} environment was highlighted as a way to enhance students' interest and participation. %Key recommendations include incorporating OSS projects into courses. % as mentioned by P57: ``\textit{It could be really cool having OSS projects directly inside courses, where I study there is nothing like this.}'' 

\textbf{Incentives and personalization.} 
%This category focuses on providing incentives and personalizing OSS projects to align with students' interests and motivations. 
Students recommended the importance of \textbf{providing financial compensation} as a motivator to contribute to OSS projects. Some recommendations include organizing paid job offerings with remuneration for contributions and pooling money for specific features. %, as stated by P87: ``\textit{Paying money (People who need a specific feature could pool money together and pay developers to implement a feature).}'' 
Other suggestions include finding ways to monetize contributions, providing payments and assistance for career benefits, and offering small financial incentives as outlined in P197: ``\textit{I think a lot of people don't like to contribute to OSS projects because they are not paid, maybe paying a little bit will help to attract more people.}''
Moreover, \textbf{rewarding contributors} in OSS projects can be an effective strategy to increase student participation and engagement. Our findings include offering academic advantages, enhancing the rewarding and fun aspects of contributing, recognizing contributors, and organizing prizes. %, as suggested in P48: ``\textit{Run something like a contest with prizes (they don't have to be monetary ones).}'' 
Hence, raising awareness about the impacts of contributing could attract more students, as outlined by P64: ``\textit{We should have more awareness about what OSS projects entail and what my contributions will bring.}''
In addition, \textbf{aligning OSS projects with personal interests} could enhance student engagement and motivation to contribute. %, as described by P183: ``\textit{An OSS project that is more in-line with what are my current interests or maybe something that could help me in the future/tools I like to use.}'' 

\textbf{Diversity and inclusion.} 
%This category addresses the need to make OSS projects more inclusive and accessible. 
Ensuring that OSS projects are \textbf{available in other languages} was mentioned as a way to enhance inclusivity and accessibility, making it easier for non-English speakers to contribute.

\rqtwo[
    \tcblower
    \textbf{Answer:} Simplifying the contribution process is the most mentioned recommendation to make OSS projects more attractive to students who have never contributed. Other recommendations include increasing awareness and offering academic benefits, job opportunities, and social influence.
]{}
%-----------------------------------

\vspace{-5px}
\section{Discussion and Implications}
\label{sec:discussion}
\vspace{-2px}

We developed a theoretical model based on OSS literature to explore the relationship between students' interest in contributing to OSS projects and various barriers and motivations. Our analysis reveals key findings and implications, which we discuss in this section. 

\textbf{Internalized Extrinsic $\rightarrow$ InterestOSS (H3).}  %Internalized extrinsic motivations encompass reputation, reciprocity, learning, and personal use. 
Our results indicated that \textit{Reputation} (path = 0.812) and \textit{Reciprocity} (path = 0.521) are associated with more interest in contributing to OSS projects. Our findings align with Lakhani and von Hippel~\cite{lakhani2004open}, who found that reciprocity motivates developers to perform routine tasks and that those who have received help from others are more likely to reciprocate as they gain experience and knowledge. Our research extends these insights by showing that reciprocity, along with reputation, not only motivates current contributors but also has the potential to attract new contributors, such as students, to join OSS projects. This finding is related to student recommendations that emerged in RQ2 (Subsec.~\ref{sec:rq2}), such as %reputation (i.e., encouraging students and creating a welcoming environment). 
highlighted in P152 ``\textit{Have incentives or certificates like how it is in hackathons.}'' (i.e., reputation).% and outlined in P153 ``\textit{Open meetings where community talk and work together.}" (i.e., reciprocity).

%This finding is related to some themes that emerged in RQ2 (Subsec.~\ref{sec:rq2}), such as: %(i) learning (i.e., mentoring, well-documented OSS projects, providing a study environment, creating training projects, and labeling tasks); 
%(i) reciprocity (i.e., providing a channel to ask questions); %(iii) own-use (i.e., aligning with personal interests); 
%and (ii) reputation (i.e., encouraging students and creating a welcoming environment). %These strategies are reflected in comments such as P66, ``\textit{Having projects or groups that focus on helping junior developers learn more and develop their skills and guide them to how that specific project works would definitely attract me.}'' and P214, ``\textit{Group projects by difficulty and have beginner-friendly projects with mentors.}''

\textbf{Intrinsic $\rightarrow$ InterestOSS (H4).} Intrinsic motivation is positively associated with an increased interest in contributing to OSS projects. Our results showed that \textit{Fun} (i.e., path = 0.919) is the indicator that has the strongest correlation with the intrinsic motivation construct. Our findings are consistent with Lakhani and Wolf~\cite{Lakhani2003Why}, who identified intrinsic motivations as the most powerful drive for individuals contributing to OSS projects. However, unlike their focus on existing contributors, our study highlights that for students--who are often outsiders to these communities--motivations play a crucial role in sparking their interest to join. 

Additionally, our findings also align with the results in RQ2 (Subsec.~\ref{sec:rq2}), where students suggested approaches linked to intrinsic motivations, such as fun. For instance, recommendations to gamify the contribution process and provide a user-friendly interface cater to the motivation of fun. %Comments like P40, ``\textit{Make levels, i.e., allow students to contribute based on their level of knowledge and understanding,}'' emphasize this aspect. Altruism is reflected in suggestions to make projects available in other languages, as highlighted by P176, ``\textit{Availability in other languages.}'' 

\textbf{Gender minorities $\times$ Extrinsic $\rightarrow$ InterestOSS (H5b).} We found support for gender minorities moderation by reducing the association between extrinsic motivational factors and the interest in contributing to OSS projects. Extrinsic motivations in OSS development impact task effort in various ways, influenced by the needs for competence, autonomy, and relatedness~\cite{ke2010effects}. 
Our findings showed significant negative moderating effects of gender on the relationship between extrinsic motivations and interest in contributing to OSS. This suggests that extrinsic motivations have less influence on minority groups compared to males, possibly due to different perceptions or the greater importance of other factors for minorities. Zaccone and Pedrini~\cite{zaccone2019effects} examined the relationship between individual motivation---both intrinsic and extrinsic---and learning effectiveness, also considering the moderating role of gender. Their findings indicate that intrinsic motivation positively affects learning effectiveness, while extrinsic motivation negatively impacts it. Similar to our study, their results show that gender plays a moderating role. In common, these results are aligned with literature suggesting gender differences in interacting with technology: women are motivated by what they can accomplish with it, while men are often motivated by their enjoyment of technology itself~\cite{burnett2010gender, margolis2002unlocking, burnett2011gender}. These gender differences can influence students' varying interests in interacting with OSS projects. %, but extrinsic motivation still positively correlates with interest in contributing. 
Furthermore, as many companies, such as Microsoft, Google, and IBM, actively hire or sponsor OSS contributors~\cite{o2021coproduction}, career ambitions and financial compensation have emerged as prevalent extrinsic motivations~\cite{schaarschmidt2018company}. \textit{Career} (i.e., path = 0.856) shows a stronger correlation than \textit{Pay} (i.e., path = 0.460). This result corroborates Jeffrey et al.~\cite{roberts2006understanding}'s findings that paid participation and status motivations in OSS developers lead to higher contribution levels, improving performance rankings. Moreover, Lin et al.~\cite{lee2017m} also report that career concern motivates individuals to contribute to online collaboration communities, signaling their qualities to potential employers.

 %Lack of diversity has significant drawbacks: (i) OSS projects miss out on the benefits of a broader range of contributors and the diverse perspectives they bring; (ii) underrepresented groups miss out on valuable learning and experience opportunities offered by these projects; and (iii) individuals from minority backgrounds may face limited job opportunities when hiring decisions use OSS contributions~\cite{marlow2013impression, santos2023designing, singer2013mutual}.

Our findings RQ2 (Subsec.~\ref{sec:rq2}) uncover strategies recommended by students. Students recommend strategies related to extrinsic motivations such as pay (i.e., providing financial compensation) and career advancement (i.e., rewarding the contributors). %, advertising the OSS projects, simplifying the contribution process, and introducing OSS at the university). 
Future work can investigate how these strategies might affect students' interest in participating in OSS. %For instance, P149 suggested, ``\textit{Count as internship. Award money. Provide job guarantees. I see no reason to contribute otherwise}'', and P48 proposed, ``\textit{Run something like a contest with prizes (they don't have to be monetary ones)}''.

In addition, Gerosa et al.~\cite{gerosa2021shifting} suggest that contributors often shift from extrinsic to intrinsic motivations over time. Therefore, they advocate that OSS communities should invest in extrinsic motivation initiatives to attract new contributors~\cite{gerosa2021shifting}. They recommend that coursework and summer coding programs, such as Google Summer of Code, can be valuable entry points for OSS involvement. However, we did not find an association between coursework and interest in contributing to OSS.

\textbf{Barriers $\rightarrow$ InterestOSS (H1).} Our analysis does not support H1, suggesting that motivations may play a more significant role in students' interest in contributing to OSS projects than the perception of barriers in OSS. This may also explain why we did not find significant support for H1, as barriers are typically encountered by those already involved in OSS rather than by potential newcomers who have yet to experience them firsthand. This implies that while challenges exist, the driving factors behind students' willingness to engage in OSS are primarily rooted in their motivations. %Moreover, we found that barriers are negatively associated with the interest in contributing to OSS projects. The barriers newcomers face can significantly impact their decision to continue contributing to an open-source software project, with \textit{Toxicity} (i.e., path = 0.451) being strongly correlated, which relates with previous literature~\cite{bosu2019diversity, guizani2021long}, which evidenced toxic situations in which OSS project members were unfriendly, unhelpful, or elitist~\cite{storey2016social}. Furthermore, the literature shows that these barriers can demotivate newcomers, often leading them to abandon their efforts~\cite{steinmacher2015social}. Our findings indicate that barriers are negatively associated with interest in contributing to open-source software projects. Newcomers frequently encounter hostile and unfamiliar environments when onboarding to an OSS project. According to Fogel~\cite{fogel2005producing}, newcomers may hesitate for a long time before attempting to contribute again if a project does not make a good first impression. This is particularly relevant for students in software engineering classes who are still developing their skills and experience. Understanding these barriers can help communities and researchers design tools and strategies to support newcomers~\cite{balali2018newcomers}. Future research should investigate the strength of the relationship between barriers and interest in contributing to OSS projects. Additionally, exploring these barriers' specific roles in reducing students' interest and identifying strategies to mitigate them could provide valuable insights.

\textbf{Involvement with OSS $\rightarrow$ Interest in OSS.} Our findings indicate that current contributors exhibit a significantly higher interest in continuing to contribute to OSS projects, which could suggest a sustained engagement among those who have already overcome initial barriers. This finding aligns with the literature that once individuals become active contributors, they develop a deeper commitment to OSS participation that could be related to having higher access privileges and higher status and peer recognition in the community, and therefore enjoy greater benefits~\cite{fang2009understanding}. Since \textit{Involvement with OSS} was positively associated with interest, it is crucial to understand what drives current contributors to remain engaged. Investigating these factors could provide insights into how to retain new contributors and help them navigate the initial challenges of OSS involvement. 

\textbf{Implications for OSS communities to attract newcomers.} Steinmacher et al.~\cite{steinmacher2014attracting} proposed a joining model in which they represent motivation and attractiveness as forces that influence outsiders to become newcomers to the OSS project and barriers as opposing forces. Even though students suggest making the contribution process as straightforward as possible, providing clear and easy-to-understand architectural documentation, reducing the intimidation newcomers feel, and making it easier for them to start contributing, our results show that motivation is a stronger predictor of interest than barriers. Therefore, highlighting the benefits of contributing, focusing on intrinsic and internalized extrinsic motivations could enhance attractiveness. 

\textbf{Implications for educators.} Familiarizing students with the OSS contribution process is becoming more common~\cite{pinto2017training}. Contributing to a real project helps students gain real-life experience and allows them to add this experience to their resume, which aids them in securing jobs. Educators can implement strategies such as gamifying the contribution process to effectively engage students by tapping into their intrinsic motivation for fun. This approach aligns with students' suggestion to ``\textit{make levels, i.e., allow students to contribute based on their level of knowledge}, which encourages active participation. 
%highlight the career benefits of contributing to OSS, such as enhancing job prospects and building a professional network to increase interest in OSS. Promoting programs like Google Summer of Code can provide valuable entry points for students. Moreover, developing beginner-friendly projects with detailed guides and mentorship can support new contributors. Grouping projects by difficulty and designing skill development projects can help students gradually build their expertise.

\textbf{Implications for research.} The significant negative moderation effect of gender on the relationship between extrinsic motivations and interest in OSS participation highlights the importance of exploring how different demographic groups are influenced by motivational factors. Future research should investigate other moderating variables, such as cultural background and educational level, to better understand what drives or hinders interest across diverse groups. Longitudinal studies could provide valuable insights into how motivations and barriers evolve over time, particularly for students, and how these factors influence long-term OSS engagement. 
%Additionally, our findings point out the importance of inclusivity, such as offering OSS in multiple languages, for attracting a diverse range of contributors. 
Future research should examine how design elements and community practices in OSS projects impact inclusivity.%, leading to best practices for creating welcoming environments for underrepresented groups.
%Future research can explore the barriers that reduce students' interest in contributing to OSS and how these barriers can be mitigated. Understanding the strength of these barriers can provide insights into designing effective support strategies. In addition, research can investigate how and why contributors' motivations shift from extrinsic to intrinsic over time. Understanding these dynamics can help OSS communities tailor their strategies to attract and retain contributors. Further studies should examine the moderating role of gender in motivation to contribute to OSS. Identifying gender-specific challenges and motivations can inform more inclusive and effective engagement strategies.

\vspace{-5px}
\section{Threats to Validity}
\label{sec:threatstovalidity}
\vspace{-2px}

\textbf{External Validity.} As detailed in Section~\ref{sec:methodology}, we recruited participants using Prolific, which may introduce inherent biases. Consequently, our conclusions are specifically valid for our sample. Future research should aim to obtain a larger sample to enhance the generalization of the findings. %Our study involved 241 students, categorized based on their experiences with contributing to OSS projects: never contributed (N=170), dropped out (N=26), and current contributors (N=45). While this segmentation may impact the generalizability of the results, our data encompasses a wide range of situations. Our findings should be regarded as a foundational starting point focusing on theoretical generalizability over statistical generalizability.

\textbf{Internal Validity.} Our model comprises four exogenous variables: barriers, extrinsic motivation, internalized extrinsic motivation, and intrinsic motivation. It also includes ``Gender'' as a moderator and ``OSSCourse'' and ``Involvement with OSS'' as control variables. While our study highlights these key factors associated with the interest in contributing to OSS, we recognize that other variables may also play significant roles. Our findings serve as a foundation for future research to explore additional associations.

\textbf{Construct Validity.} We adapted and customized existing measurement instruments for various constructs based on established OSS literature. Additionally, we piloted the study to gather feedback on our instrument, mitigating potential threats to its validity. These iterative pilots allowed us to refine and validate our survey, ensuring its effectiveness and reliability in accurately capturing the intended constructs. Furthermore, our analysis of the measurement model verified that these constructs demonstrated internal consistency and achieved satisfactory results in both convergent and discriminant validity tests.

\textbf{Subjectivity.} We employed qualitative procedures to classify responses to the open-ended questions, which are inherently subject to interpretation bias. We adopted a multi-faceted approach involving collaboration among multiple researchers to mitigate this potential bias. The team engaged in continuous comparative analysis and reached conclusions through a process of negotiated agreement. Each team member brings extensive experience in qualitative methods and OSS, enhancing the robustness of our analysis.

\section{Related Work}
\label{sec:relatedwork}
\vspace{-2px}

In this section, we present studies related to newcomers, motivations to contribute, and students' perceptions of OSS projects.

\textbf{Newcomers in OSS.} Newcomers face several barriers in OSS~\cite{steinmacher2015systematic, steinmacher2015understanding, santos2022hits}, which may lead to high dropout rates~\cite{steinmacher2013newcomers}. Previous studies have examined the process of newcomers joining community-based OSS projects, offering valuable insights into the factors influencing their experiences within OSS communities~\cite{PS02canfora2012going, park2009beyond, steinmacher2013newcomers, steinmacher2014attracting, PS18wang2011bug} and identified software solutions that facilitate the onboarding of newcomers in software projects~\cite{santoslr}. For example, factors such as project popularity, review time for pull requests, project age, programming languages, and problem-solving styles affect the onboarding of new contributors in OSS projects~\cite{fronchetti2019attracts, santos2023designing}. Understanding these factors helps project maintainers optimize their strategies for onboarding new contributors. The project type is a factor that impacts underrepresented groups. Women are more likely to select OSS projects for social good (OSS4SG) than men~\cite{fang2023four}, and students contributing to OSS4SG projects are significantly more likely to have contributions accepted by their communities. Our study stands out from existing literature by focusing on developing a theoretical model that elucidates the relationship between students' perceptions of OSS and their interest in contributing to OSS projects.

\textbf{Motivations to contribute to OSS projects.} According to Gerosa et al.~\cite{gerosa2021shifting}, motivation to contribute to OSS was extensively studied. Considerable research has been dedicated to understanding motivations for joining OSS~\cite{roberts2006understanding, von2012carrots, alexander2002working}. These studies examined specific communities~\cite{choi2015characteristics, spaeth2015research} and various contributor profiles, including newcomers~\cite{hannebauer2017relationship}, one-time code contributors~\cite{lee2017understanding}, and students~\cite{silva2020google}. Prior research indicates multiple factors influence an OSS contributor's decision to join, remain, or leave a project~\cite{steinmacher2014attracting, kaur2022exploring}. 

\textbf{Students perceptions.} Holmes et al.~\cite{holmes2018dimensions} explored students' perceptions of contributing to OSS, finding that students valued the opportunity to apply their skills to real-world tasks and receive authentic feedback from project maintainers. Similarly, Pinto et al.~\cite{pinto2019training} examined students' views on the requirement to contribute to an OSS project as part of a Software Engineering course, offering insights into how these experiences shape their learning and engagement. Differently from previous work, we have investigated the factors associated with students' interest in contributing to OSS projects. Understanding these factors can lead to developing strategies to attract students to the OSS workforce.

\section{Conclusion}
\label{sec:conclusion}
\vspace{-2px}

%We surveyed 241 students to develop a theoretical model to understand the link between students' perceptions of OSS and their interest in contributing to it. By examining students' perceptions, we can help communities strategize effectively to attract newcomers.

Our study identified several key factors driving students' interest in OSS. Intrinsic motivation emerged as a strong influencer. Internalized extrinsic motivation also played a significant role. Interestingly, the impact of extrinsic motivation varied by gender, positively affecting males but negatively influencing minority groups. Our model can aid researchers and community leaders in designing targeted interventions by highlighting the factors influencing students' interest in OSS. Furthermore, students provided recommendations to enhance the attractiveness of OSS projects. They emphasized the need to simplify the contribution process, increase awareness, and offer academic and job-related benefits. Other strategies include providing mentoring, rewarding contributors, creating a welcoming environment, and aligning projects with personal interests. In future work, we aim to design interventions that make OSS projects more attractive to students and developers in general.

\section*{Acknowledgments}
\vspace{-2px}

The National Science Foundation (NSF) partially supports this work under grant numbers 2247929 and 2303042. Katia Felizardo is partially funded by a research grant from the Brazilian National Council for Scientific and Technological Development (CNPq), Grant 302339/2022--1.

\bibliographystyle{ACM-Reference-Format}
\bibliography{bibtex.bib}

\end{document}